\documentclass[manuscript]{aastex}

\shorttitle{Magnetic outbreak}

\shortauthors{Jin et al.}

\begin{document}


\title{Magnetic outbreak associated with exploding granulations}


\author{Chunlan Jin and Guiping Zhou}
\affil{Key Laboratory of Solar Activity and Space Weather, National
Astronomical Observatories, Chinese Academy of Science, Beijing
100101, China; cljin@nao.cas.cn}

\author{Guiping Ruan}
\affil{Shandong Provincial Key Laboratory of Optical Astronomy and
Solar-Terrestrial Environment, and Institute of Space Sciences,
Shandong University, Weihai 264209, China}

\author{T. Baildon and Wenda Cao}
\affil{Center for Solar-Terrestrial Research, New Jersey Institute
of Technology, 323 Martin Luther King Blvd, Newark, NJ 07102,
U.S.A.}
\affil{Big Bear Solar Observatory, 40386 North Shore Lane,
Big Bear City, CA 92314, U.S.A.}

\and

\author{Jingxiu Wang}
\affil{Key Laboratory of Solar Activity and Space Weather, National
Astronomical Observatories, Chinese Academy of Science, Beijing
100101, China} \affil{School of Astronomy and Space Science,
University of Chinese Academy of Sciences, 100049, Beijing, China}



\begin{abstract}
Diagnosing the spatial-temporal pattern of magnetic flux on the Sun
is vital for understanding the origin of solar magnetism and
activity. Here, we report a new form of flux appearance, magnetic
outbreak, using observations with an extremely high spatial
resolution of 0.16 arcsec from the 1.6-m Goode Solar Telescope (GST)
at the Big Bear Solar Observatory. Magnetic outbreak refers to an
early growth of unipolar magnetic flux and its later explosion into
fragments, in association with plasma upflow and exploding
granulations; each individual fragment has flux of
10$^{16}$-10$^{17}$ Mx, moving apart with velocity of 0.5-2.2 km/s.
The magnetic outbreak takes place in the hecto-Gauss region of pore
moats. In this study, we identify six events of magnetic outbreak
during 6-hour observations over an approximate 40$\times$40
arcsec$^{2}$ field of view. The newly discovered magnetic outbreak
might be the first evidence of the long-anticipated convective
blowup.

\end{abstract}





\section{Introduction}

The appearance of magnetic flux on the solar surface manifests a
fundamental process that energizes solar atmosphere and leads to
solar eruptions(e.g., Magara \& Longcope 2003; Guglielmino et al.
2010). Flux appearance signifies the physical interaction between
plasma motion and generated magnetic fields. For the past century,
observations have shown that magnetic flux emerges in a bipolar form
on the Sun, from the strong field regime, e.g., in active regions,
to the weak field regime and even in internetwork regions (Schrijver
\& Zwaan 2000; Cheung et al. 2010; Stein et al. 2011; Wang et al.
2012). However, an exception to this is moving magnetic features
(MMFs), which have drawn intense attention in the solar physics
community (Sheeley 1969; Harvey \& Harvey 1973); MMFs do not
typically exhibit the evolutionary pattern of an emerging flux
region, i.e., the systematic growth and separation of opposite
polarities, although Type I MMFs do appear in bipoles (Wilson 1986;
Spruit et al. 1987; Thomas et al. 2002; Zhang et al. 2003). Type II
and III MMFs, however, are unipolar magnetic features (Shine \&
Title 2000).

In the past years, a number of small-scale flux emergence events
occurring at mesogranular scale and granular scale have been
studied. Observations show that the exploding granule (EG) is
associated with the flux emergence occurring at mesogranular scale
(Goglielmino et al. 2020), and contributes to organize the discrete
magnetic field (e.g., Roudier et al. 2016; Malherbe et al. 2018;
Roudier et al. 2020). Granule-covering magnetic sheet-like
structures in the quiet Sun have been found by the observations
(Centeno et al. 2017; Fischer et al. 2019) and the numerical
simulations (Moreno-Insertis et al. 2018). Furthermore, the
appearance of unipolar features in internetwork (IN) flux has been
observed (Go$\breve{s}$i$\acute{c}$ et al. 2022).

Based on magnetic observations with extremely high-spatial
resolution, we find a new magnetic phenomenon with unipolar form:
magnetic outbreak. This phenomenon is found in pore moats - the same
magnetic environment as MMFs. In this paper, we will present our new
findings in detail. Observations and data analysis are presented in
Section 2, and in Section 3, we give a detailed description of the
magnetic outbreak phenomenon. We discuss our revelation in the
context of previous findings and provide some possibilities to
explain the new observation in Section 4, and conclusions are made
in Section 5.

\section{Observations and data analysis}
Extremely high spatial resolution observations of NOAA active region
(AR) 12579 on 25 August 2016 were achieved by the 1.6-m Goode Solar
Telescope (GST; Goode \& Cao 2012; Cao et al. 2010). The
observations were made with the Near InfraRed Imaging
Spectropolarimeter (NIRIS; Cao et al. 2012, 2022) over the 1.56 $\mu
m$ Fe I line at the Big Bear Solar Observatory (BBSO), and have high
spatio-temporal resolution: approximate 57 km/pixel and 41s cadence.
NIRIS produces full spectropolarimetric measurements $I$, $Q$, $U$,
and $V$ (Stokes profiles) at a spectral resolution of 0.01 nm, with
a typical range of -0.32 nm to +0.31 nm from the line center.
Broadband TiO images centralized at 705.7 nm were obtained with a
high spatial resolution of 25 km/pixel and temporal resolution of 15
s at BBSO. Images and magnetograms from the Solar Dynamics
Observatory (SDO; Pesnell 2012) were also used for coordinative data
analysis.

Each NIRIS data sample (pixel) is comprised of Stokes profiles taken
at 55 spectral points. The rms fluctuation of the spectral continuum
is 0.11\% in the Stokes $Q$ and $U$ spectra, and 0.09\% in the
Stokes $V$ spectrum. The NIRIS data undergoes Stokes inversion based
on the Milne-Eddington atmospheric model (Ahn et al. 2016), through
which several physical parameters, including vector magnetic field
and Doppler shift, have been obtained. The magnetic signal as low as
4 G can be detected for the line-of-sight field. The accuracy of the
resulted vector field data reaches 10 G for line-of-sight component
and 100 G for transverse component (Wang et al. 2017). In addition,
the inverted Doppler velocity is calibrated by setting the average
Doppler velocity of the very quiet region to be zero.

\begin{figure}
\includegraphics[angle=0,width=0.8\textwidth]{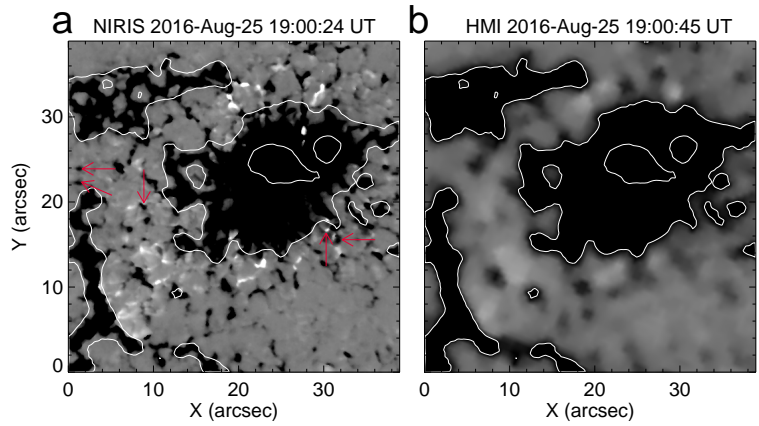}
\caption{Comparison of magnetic observations between
NIRIS and HMI. The line-of-sight magnetic observations from NIRIS
(a) and HMI (b) were taken on 25 August 2016 at 19:00 UT. The
magnetic field saturates at $\pm$100 G. Large-scale structures are
outlined in white in both images. \label{fig1}}
\end{figure}

The NIRIS line-of-sight magnetic observations are compared with
magnetic measurements from the Helioseismic and Magnetic Imager
(HMI; Scherrer et al. 2012) instrument onboard SDO. HMI enables
magnetic observations with a pixel size of 362 km and cadence of 45
s. The compared magnetograms are shown in Fig.1. We can see similarity in the larger-scale
magnetic structures between the NIRIS and HMI magnetograms, and more
fine-scale structures in NIRIS magnetic observations due to higher
spatial resolution. We further compare the magnetic flux of the main
magnetic structures, which are outlined in white in Fig.1, and
obtain a flux measurement ratio of 1.2 between NIRIS and HMI. A few
magnetic elements indicated by arrows in the NIRIS magnetogram are
completely missing in the HMI magnetogram. This illustrates the
advantage of using NIRIS observations for exploring small-scale
magnetic evolution on the solar surface (Wang et al. 2017), and
small-scale magnetic structures can only be revealed via high
spatial resolution (Jin \& Wang 2019).

\section{Magnetic outbreak phenomenon}

\begin{figure}
\includegraphics[angle=0,width=1.\textwidth]{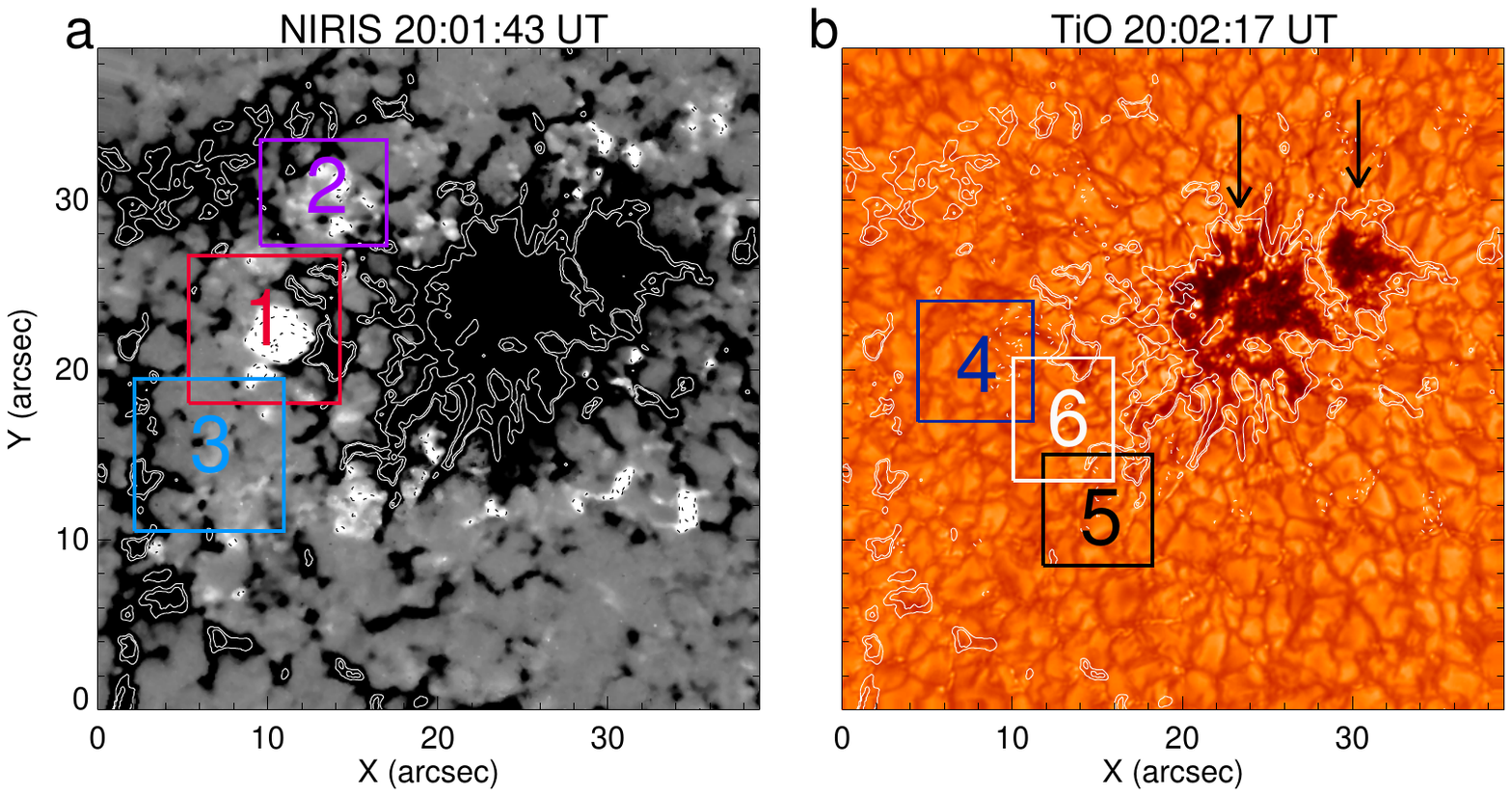}
\caption{The magnetic environment of magnetic outbreak events. The
positions of outbreak events are indicated in the NIRIS magnetograms
(scaled between $\pm$ 50 G) (a) and TiO image (b). The six events
appear in the moats of pores. The solid and dotted contour lines
correspond to negative and positive magnetic field values of [-1000
G, -300 G, 50 G, 200 G]. \label{fig2}}
\end{figure}

Approximately 6-hour continuous observations from 16:36 UT to 22:16
UT were taken of the negative polarity region of AR 12579. The
negative magnetic region mainly consists of two pores indicated by
the black arrows in Fig. 2b. We identify six events of magnetic
outbreak during the observations; the locations of these events are
distributed in the moats of pores, which are framed in Fig.2a and
2b. The primary properties of magnetic outbreak are listed in Table
1. In this table, the foreshortening effect may affect the values of
line-of-sight magnetic component, because the observation is not
acquired at disk center. However, considering the quieter magnetic
environment of these outbreak events and the larger errors from the
transverse field, the foreshortening effect is not corrected in this
study.

\begin{table}
\begin{center}
\caption{Fundamental properties of the six observed magnetic
outbreak events. \label{tbl-1}}
\begin{tabular}{crrrrrrrrrrr}
\tableline\tableline Event & Appearing & Maxflux & Ending &
Lifetime & Flux & LOS & Transverse& Velocity\tablenotemark{b} \\
No. & time(UT) & time(UT) & time(UT) &
 (min) & (Mx) & field\tablenotemark{a}(G) &  field\tablenotemark{a}(G) &  (km s$^{-1}$) \\
\tableline
1 &19:50 &20:04 &21:16       &86 &1.1e19  &107 &317  &1.3/0.7 \\
2 &19:30 &19:43 &20:27       &57 &3.0e18  &85  &245  &0.6/0.8 \\
3 &21:27 &21:54 &after 22:16 &   &1.2e19  &60  &172  &1.2/0.5\\
4 &17:58 &18:21 &18:57       &64 &5.3e18  &68  &180  &1.1/0.5 \\
5 &19:43 &19:48 &20:15       &32 &1.4e18  &40  &151  &2.2/0.6 \\
6 &20:41 &21:00 &21:58       &77 &3.4e18  &50  &199  &0.5/0.9 \\
\tableline
\end{tabular}
\tablenotetext{a}{Both LOS field and transverse field mean the
magnetic flux density of outbreak events.} \tablenotetext{b}{The
velocity is obtained by two methods, i.e., the time-slit of
magnetogram and FLCT. A continuous observation, during which the
exploding phase occurs, is adopted to obtain the velocity based on
the FLCT method. The used FOVs for FLCT are labeled by square frames
in Fig.2 for the six outbreak events.}

\end{center}
\end{table}

\begin{figure}
\includegraphics[angle=0,width=1 \textwidth]{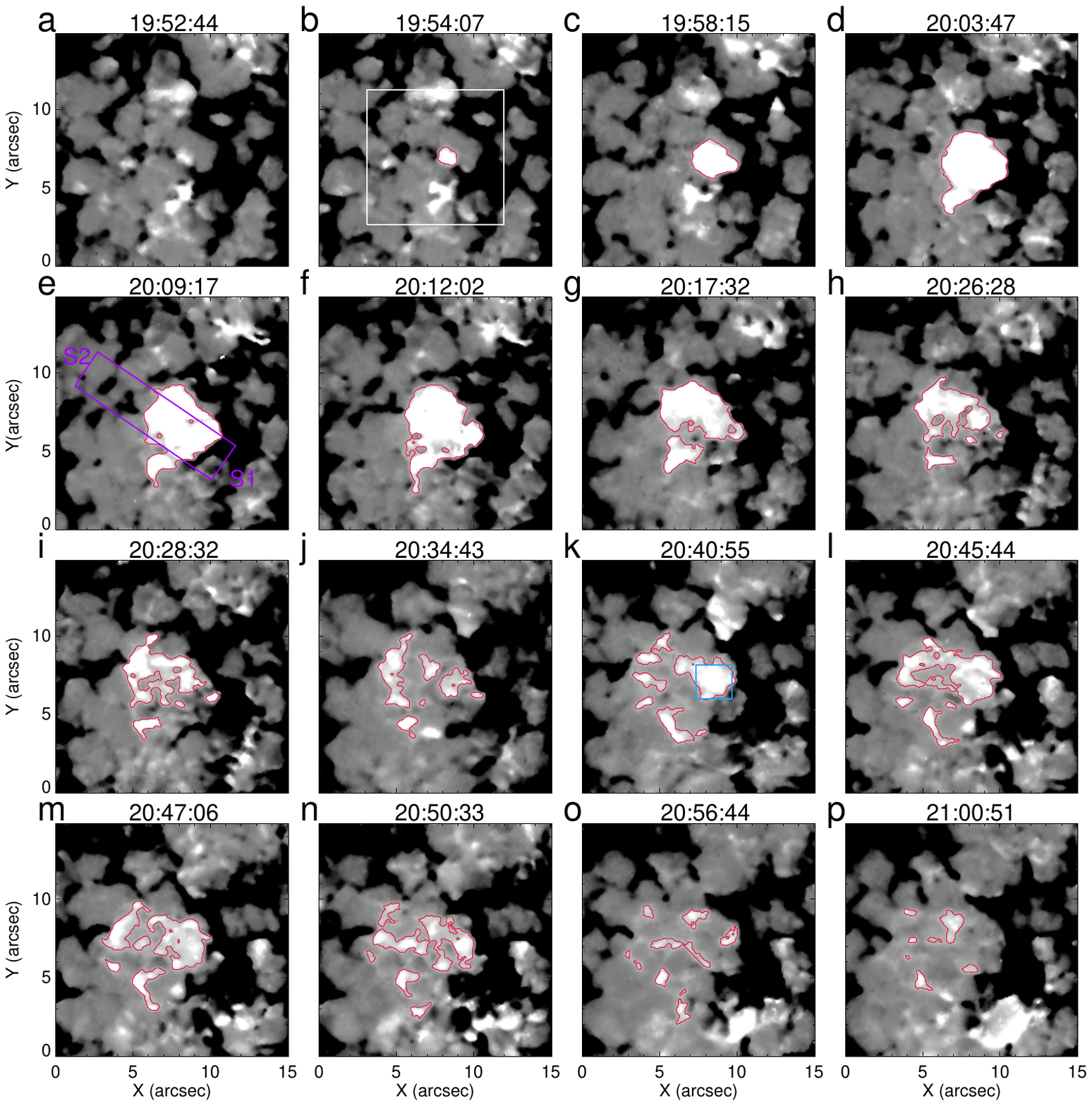}
\caption{Time series of NIRIS magnetograms (scaled between $\pm$ 50
G) showing the process of outbreak Event 1. The outbreak flux
patches are outlined in red. The magnetic field framed in blue in
(k) indicates the secondary outbreak. The slit S1-S2 in (e) is used
to obtained the velocity of magnetic outbreak patches. The area
framed in white box in (b) has the same field of view as the red box
in (a) of Fig.2. An animation of this figure is available. The
animation lasts 4 s and covers 1.45 hr of solar time from 2016
August 25 at 19:48 UT. As a comparison, the corresponding HMI
magnetic evolution is also available in the animation. \label{fig3}}
\end{figure}

\begin{figure}
\includegraphics[angle=0,width=1.\textwidth]{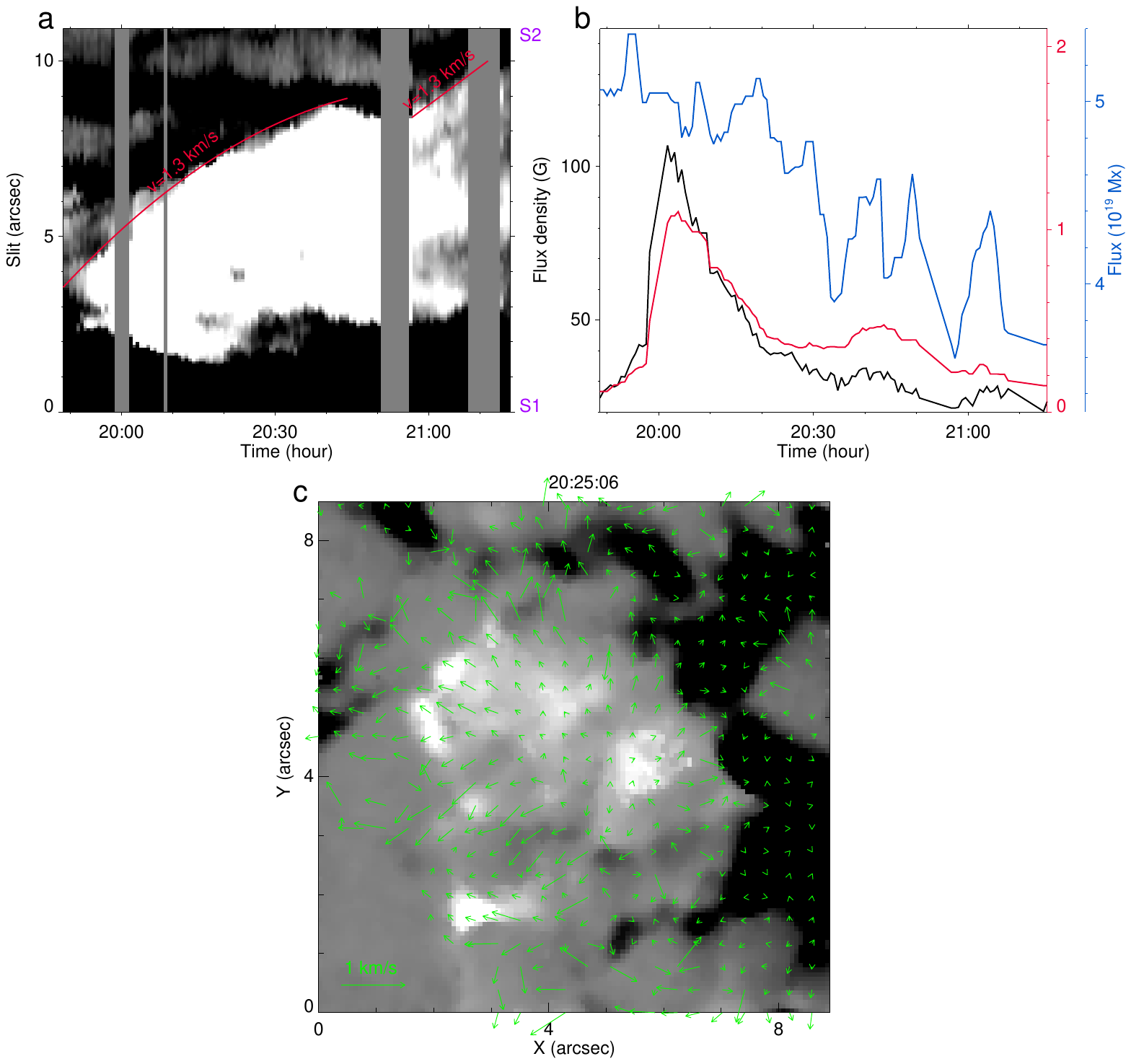}
\caption{Top panels: The time slit and magnetic variation for Event
1. (a) The time series of the slit in Fig. 3e. This data is used to
estimate the exploding velocity. Grey regions denote missing
observations. An average velocity of 1.3 km/s is obtained by the
fitting. (b) Magnetic variations in the domain framed in white in
Fig. 3b are plotted for the flux density (black line) and flux (red
line) of the magnetic outbreak, as well as the magnetic flux of the
negative field (blue line). Bottom panel: The velocity distribution
for Event 1 ranging from 20:10 UT to 20:50 UT, which is obtained by
the FLCT method in a ~9"$\times$9" subFOV. Event 1 is centered on
(5",5"), and its average velocity is 0.7 km/s. \label{fig4}}
\end{figure}

\begin{figure}
\includegraphics[angle=0,width=0.8 \textwidth]{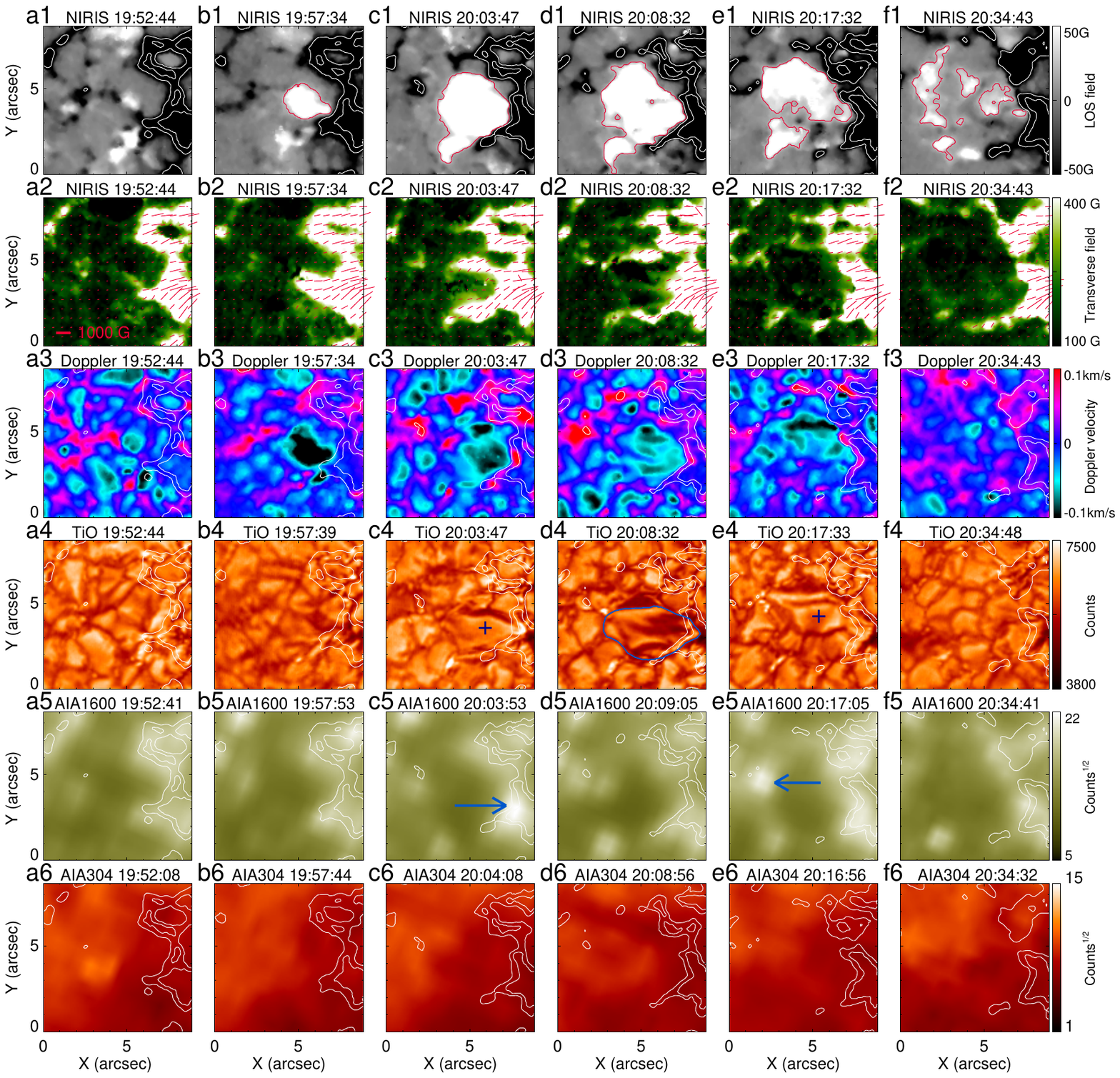}
\caption{Time series images showing the convection and atmospheric
response during Event 1. First row: line-of-sight magnetograms.
Second row: transverse magnetograms, where the transverse field with
the value larger than 100 G are shown by red lines. Third row:
Dopplergram. Fourth row: TiO images. Fifth and Sixth rows: AIA 1600
{\AA} and 304 {\AA} images. Contour lines of [-1000 G, -100 G] from
synchronous NIRIS line-of-sight magnetograms are marked by the white
lines in these images. The field-of-view of these images is the same
as that in red in Fig.2. An animation of this figure (including
NIRIS LOS field, Doppler velocity, AIA1600 and AIA304 images) is
available, and these images in animation are shown by a larger field
of view, which is the same as that in Fig.3. The animation lasts 4 s
and covers 1.45 hr of solar time from 2016 August 25 at 19:48
UT.\label{fig5}}
\end{figure}

Event 1 (labeled "1" in Fig.2) is a typical event of magnetic
outbreak with relatively longer duration and larger magnetic flux,
and thus, is worthy of a detailed discussion. Event 1 takes place in
a cell region of mesogranular size (November et al. 1981). The
process of magnetic outbreak consists of three phases: the growing
phase, the exploding phase, and the fading phase, all of which are
shown in Fig.3. The detailed evolution within the three phases is
described below.
\begin{itemize}
\item
\textbf{The growing phase.} Starting around 19:50 UT, a lump of
positive magnetic flux appears and grows, reaching a maximum flux
density of 107 G at about 20:02 UT. This growing phase is
characterized by the rapid increase of both total flux and flux
density (see Fig.4). In this period, the positive flux keeps its
center position unchanged (see Fig.3), while its area expanded. The
flux gradually increases to 1.1 $\times$ 10$^{19}$ Mx at about 20:04
UT; meanwhile there is no corresponding increase of negative flux
(see Fig.4). Furthermore, the transverse field emerges and develops
in this period, and its density reaches 317 G around 20:02 UT, which
reveals a highly inclined field of the event (see Fig.5). The growth
of positive flux is accompanied by concentrated blue shifts in NIRIS
Dopplergrams, indicating an updraft of -0.1 km/s (see Fig.5).
Moreover, an EG (Musman 1972; Rempel 2018; Roudier et al. 2020;
Guglielmino et al. 2020) appears around 20:00 UT, right after the
early flux growth. The EG grows to mesogranulation-scale at
approximate 20:08 UT; another EG then appears in the dark notch of
the former (shown by the '+' in Fig.5). In this growth period,
transient brightening is found in the upper photosphere, but there
is no obvious response in the chromosphere (see Fig.5).
\item
\textbf{The exploding phase.} Around 20:12 UT, the positive flux
begins to split into many fragments which move apart with an average
velocity of 1.3 km/s, looking like an exploding bomb (see Fig.3).
Furthermore, the velocity distribution acquired by the Fourier local
correlation tracking (FLCT: Fisher \& Welsch 2008) method also
displays the exploding property, which is shown in the bottom panel
of Fig.4. Here, the width $\sigma$ of the Gaussian windowing
function for FLCT method adopts 15 pixels size, i.e, 1.2 arcsec. At
about 20:34 UT, the main explosion completes, and the positive flux
fragments can be seen marking the explosion fronts, while we also
see ordinary granulations and calming of plasma upflows (see Fig.5).
In this process, the flux density and magnetic flux of the positive
flux gradually decreases (see Fig.4). Interestingly, a secondary or
subsequent outbreak is observed in a subset window (framed in
Fig.3k), and its explosion fronts are outlined by exploding
fragments of the positive flux. For this secondary outbreak, the
total flux reaches a maximum of 1.5$\times$ 10$^{18}$ Mx at about
20:43 UT (see Fig.4). The same type of secondary magnetic outbreak
is also observed in Event 2.
\item
\textbf{The fading phase.} The process lasts until 21:16 UT, when
all the flux fragments in the primary and secondary outbreaks
disappear. These explosion fragments either cancel with the
surrounding negative field or diffuse to levels below the magnetic
noise.
\end{itemize}
We refer to the whole process - from the first appearance of
unipolar positive flux to the disappearance of the exploding
fragments - as "magnetic outbreak" in this study. We note that the
ambient enhanced network is forced to decay and move farther from
the pores during magnetic outbreak (see Fig.3).

Only for the third among the six events, observations are truncated
due to bad weather during the late fading phase. All of the magnetic
outbreak events take place in the moats of pores, and they share a
few key properties: 1) appearance of unipolar positive magnetic flux
with hecto-G flux density, which is accompanied by plasma upflow; 2)
increase in both magnetic flux and flux density during the positive
flux growth, without in-phase changes of negative flux; 3) eruption
of EGs during the flux growing and exploding phases; and 4) weak
transient brightening appearing at the border between the positive
flux and the enhanced negative network in the upper photosphere,
without chromospheric correspondence in radiation.


\section{Discussion}

\subsection{The resemblances and differences of magnetic outbreak with previous findings}

Magnetic outbreak displays many similarities with MMFs, in view of
its relation to pore/spot moats, spatial size, flux level, and
moving velocity. It is worth noting that the non-uniform magnetic
explosion results in an apparent outflow of flux elements in the
moat, which can be seen in lower resolution observations. We see
Event 1 in the time sequence of 1$\sim$2 arcsec resolution HMI
magnetograms (see Fig.3 Movie), which looks like Type III MMFs,
i.e., outflowing magnetic features with polarity opposite to the
parent pore. On the other hand, in the GST/NIRIS magnetograms many
Type III MMFs are observed, but they never exhibit an exploding
nature (Li et al. 2019). Therefore, magnetic outbreak seems to unfit
the scenario of MMFs.

We carefully checked the magnetic and velocity observations, and
considered all the possibilities and, in particular, whether or not
the observations fit in an already known 'family' in published
literature, e.g., Moreno-Insertis et al. (2018), Guglielmino et al.
(2020), Roudier et al. (2020), etc. The answer is not exactly.
First, the magnetic outbreak is not a phenomenon within a granule,
but takes place in the interior of the enhanced network at
mesogranular scale. Plasma up-flow plays a role in triggering the
EGs and the later magnetic explosion (see the Dopplegrams and
granule images in Fig.5), similar to that described by Guglielmino
et al. (2020). Secondly, the magnetic outbreak is not related to the
IN horizontal elements (Lites et al. 2008). Jin et al. (2009) is the
first to classify the IN horizontal elements into two classes - one
class is associated to a pair of line-of-sight elements,
representing a small-scale loop emergence; the other is isolated
from line-of-sight elements. However, their studied horizontal IN
elements does not show eruptive behavior. Thirdly, the field
configuration of magnetic outbreak is not like the horizontal flux
sheets covering a whole granule by the simulation (Moreno-Insertis
et al. 2018) and the horizontal sheet emergence followed by
basically the bipolar appearance in the observation (Fischer et al.
2019). The well-organized inclined fields in outbreak events are
manifested by prevailing positive line-of-sight flux and stronger
transverse field with connection to the surrounding negative
network. Fourthly, our observations demonstrate a convective
instability initiated with updraft of plasma, which leads to the
explosion in the studied magnetic outbreak. In these events, the
enhanced network might play some role in penetrating radiation to
heat its interior which was initiating the instability. To our
knowledge, the convective instability triggered by plasma upward
motion has not been previously described in the existing literature.

\subsection{How do we understand the observed magnetic outbreak?}

\begin{figure}
\begin{center}
\includegraphics[angle=0,width=0.8 \textwidth]{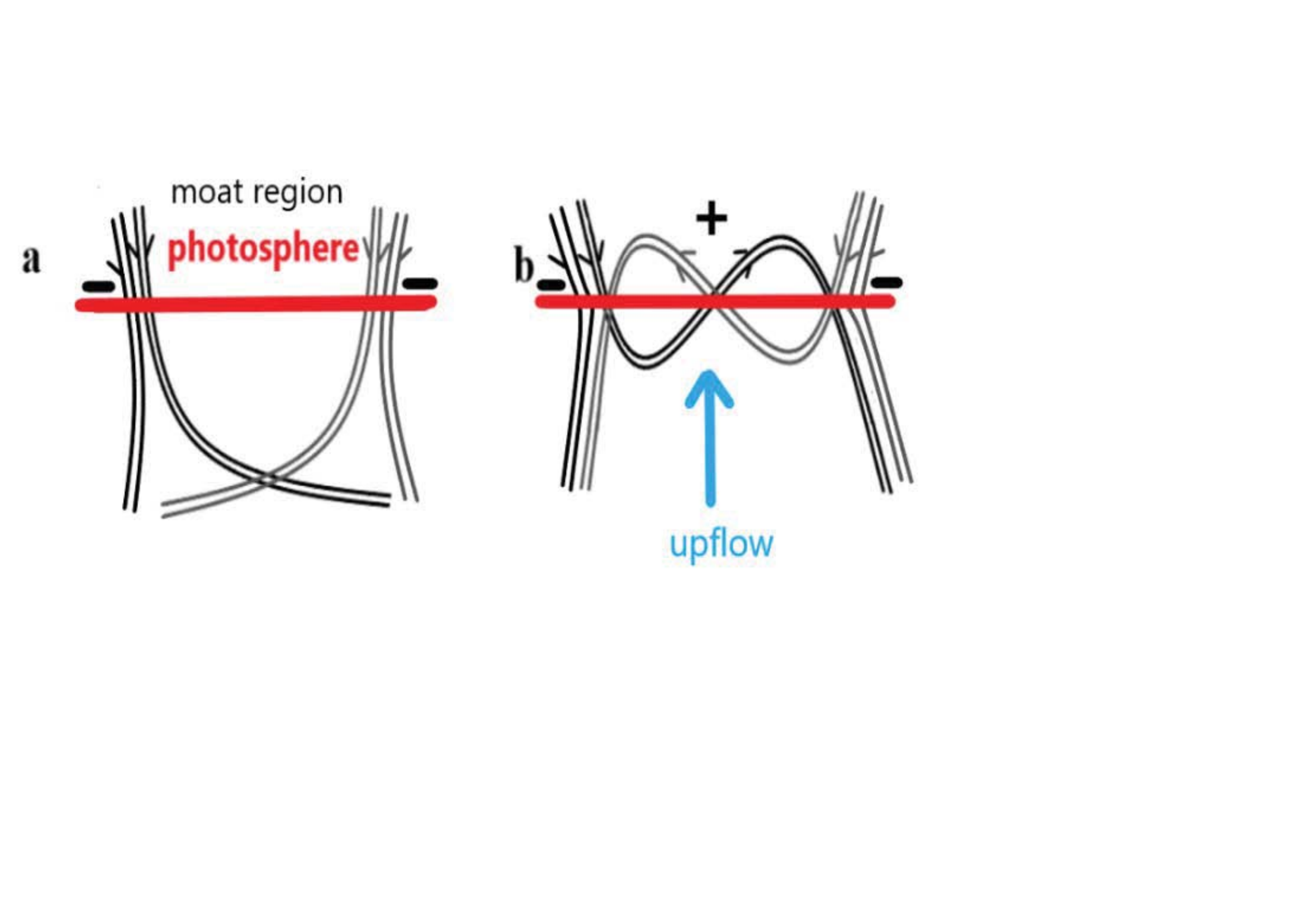}
\caption{Sketch that may account for the unipolar flux in the
observed magnetic outbreak events: Drafting by plasma upflow. Red
lines indicate the photospheric layer, and the region between the
black and grey magnetic lines is the pore moat. \label{fig6}}
\end{center}
\end{figure}

The observed magnetic outbreak raises several interesting questions,
such as why we only see the growth of uniform positive flux, what is
the basic magnetic topology which serves as the magnetic outbreak,
what physics stands behind the observations?

We tentatively propose the magnetic topology gestated the magnetic
outbreak, which could be simplified as a U-loop with an open bottom
as seen in the sketch in Fig.6. The strong vertical field at the
network boundary likely connects a reservoir of horizontal field
underneath the solar surface at a depth between the bottoms of
mesogranulation and supergranulation(Rempel \& Cheung 2014; Cheung
et al. 2007). The overall magnetic configuration is quite similar to
U-loops in the literature (Wilson 1989), except for its open bottom.
Penetration from lateral radiation would result in the temperature
rise, causing the plasma to flow upward (see Schrijver \& Zwaan
2000), which stretches the horizontal magnetic strands. As a net
result of many such up-stretched field lines, we would see the
appearance of equivalent unipolar flux with positive polarity.
Moreover, by the tension force of stretched field, the emerged
magnetic flux should be strongly inclined. In other words, one would
see the inclined flux with unipolar positive flux and strong
horizontal field component connecting to the negative network.

In contrast to the well-known convective collapse, as soon as a
convective instability occurs in the upflow with the field, the
initial updraft would be enhanced and the flux tube expands. As a
result of this increased instability, the flux tube would be torn to
shreds and the plasma returns to a normal convective state (Spruit
\& Zweibel 1979; Spruit 1979; Schrijver \& Zwaan 2000). The
convective blowup of magnetic tube was predicted more than 40 years,
but it has never been observed. Observations with the extremely high
spatial resolution and good polarization sensitivity from GST might
enabled us to report the first apparent evidence of convective
blowup.

The observed magnetic outbreak has vividly illustrated the generally
physical picture of convective blowup. The most striking
characteristics in the observed magnetic outbreak are the coinciding
plasma upflow and flux explosion. We seem to witness the convective
instability in plasma upflow and the real-time blowup of magnetic
flux. In addition, the observations have shown how the rapid
development of plasma upflow results in EGs. The EGs appear to be
involved in convective blowup. It is convective instability that
conducts the magnetic outbreak.

The identified magnetic outbreak in this study takes place in the
fading phase of AR 12579. The six outbreak events have a total flux
of about 3.6$\times$10$^{19}$ Mx, which is approximately 11\% of the
total flux loss in the parent pore region. This is indicative that
convective blowup plays a role in the removal of magnetic flux from
the pore region. We consider the conjugated convective collapse and
blowup to play key role in shaping the spatiotemporal pattern,
followed by vigorous flux emergence and cancelation. The former
creates various strong field structures, and the latter transforms
the magnetic flux from the strong field realm to the weak field
reservoir.

\section{Conclusion}
Based on high spatial resolution observations from the 1.6-m GST at
BBSO, we find a new form of flux appearance, i.e., magnetic
outbreak, in the hecto-Gauss region of pore moats. Rapid emergence,
explosion, and final dissipation constitutes the whole process of
magnetic outbreak. Magnetic outbreak is associated with plasma
upflows and EGs, and results in weak transient brightening in the
upper photosphere, without chromospheric correspondence in
radiation. During 6-hour observations, six events of magnetic
outbreak were identified in an approximate 40 arcsec field of view
in the negative polarity region of AR 12579; their magnetic fluxes
ranged from 10$^{18}$ Mx to 10$^{19}$ Mx, and their lifetime was
around an hour. The velocity of their exploding fragments reached
2.2 km/s. The newly-observed magnetic outbreak vividly describes the
physical picture of convective blowup of flux tube, and might
provides the first evidence of the long-expected convective blowup.

\acknowledgments

This work was supported by National Key R\&D Program of China
No.2021YFA1600500, the B-type Strategic Priority Program of the
Chinese Academy of Sciences (Grant No. XDB41000000), Key Research
Program of Frontier Sciences, CSA (Grant No. ZDBS-LY-SLH013), Yunnan
Academician Workstation of Wang Jingxiu (No. 202005AF150025), and
the National Natural Science Foundation of China (12273061,
11873059, and 11533008). T. Baildon and W. Cao acknowledge the
support from US NSF grants - AGS-1821294 and AST-2108235. We are
grateful to Dr. Jack Harvey for helpful discussions and valuable
comments. We would like to thank the teams of BBSO and SDO in
obtaining the data. BBSO operation is supported by NJIT and US NSF
AGS-1821294 grant. GST operation is partly supported by the Korea
Astronomy and Space Science Institute and Seoul National University.
We particularly thank the anonymous referee for the valuable
comments and helpful suggestions.

\end{document}